\documentclass[fleqn,usenatbib]{mnras}

\usepackage{newtxtext,newtxmath}

\usepackage[T1]{fontenc}

\DeclareRobustCommand{\VAN}[3]{#2}
\let\VANthebibliography\thebibliography
\def\thebibliography{\DeclareRobustCommand{\VAN}[3]{##3}\VANthebibliography}

\usepackage{graphicx}	
\usepackage{amsmath}	
\usepackage{dirtytalk} 
\usepackage{csvsimple}  

\hypersetup{
	colorlinks=true,       
	linkcolor=blue,        
	citecolor=blue,        
	filecolor=magenta,     
	urlcolor=cyan         
}
\usepackage{siunitx}
\DeclareSIUnit \h {\ensuremath{\mathit{h}}}
\DeclareSIUnit \pc {pc}

\title{The impact of weak lensing on Type Ia supernovae luminosity distances}

\author[P.Shah et al.]{
Paul Shah,$^{1}$\thanks{E-mail: paul.shah.19@ucl.ac.uk}
Pablo Lemos,$^{2}$
Ofer Lahav$^{1}$
\\
$^{1}$Department of Physics and Astronomy, University College London, Gower Street, London, WC1E 6BT, UK\\
$^{2}$Mila, 6666 St-Urbain Street, 200 Montreal, QC, H2S 3H1, Canada\\
}
\date{Accepted XXX. Received YYY; in original form ZZZ}

\pubyear{2022}

\begin{document}
\label{firstpage}
\pagerange{\pageref{firstpage}--\pageref{lastpage}}
\maketitle

\begin{abstract}
When Type Ia supernovae are used to infer cosmological parameters, their luminosities are compared to those from a homogeneous cosmology. In this note we propose a test to examine to what degree SN Ia have been observed on lines of sight where the average matter density is \textit{not} representative of the homogeneous background. We apply our test to the Pantheon SN Ia compilation, and find two redshift bins which indicate a moderate bias to over-density at $\sim 2\sigma$. We modify the Tripp estimator to explicitly de-lens SN Ia magnitudes, and show that this reduces scatter of Hubble diagram residuals. Using our revised Tripp estimator, the effect on cosmological parameters from Pantheon in $\Lambda$CDM is however small with a change in mean value from $\Omega_{\rm m} = 0.317 \pm 0.027$ (baseline) to $\Omega_{\rm m} = 0.312 \pm 0.025$ (de-lensed). For the Flat $w$CDM case it is $\Omega_{\rm m} = 0.332 \pm 0.049$ and $w = -1.16 \pm 0.16$ (baseline) versus $\Omega_{\rm m} = 0.316 \pm 0.048$ and $w = -1.12 \pm 0.15$ (de-lensed). We note that the effect of lensing on cosmological parameters may be larger for future high-z surveys. 

\end{abstract}





\section{Introduction}
\label{sec:intro}

Type Ia supernovae are used extensively in cosmology as, once standardised, their absolute magnitudes have a low and well-characterised scatter. As they are observed from our cosmic neighbourhood out to redshift $z \sim 2$, they may be used to build a Hubble diagram of their apparent magnitudes and redshift which connects the expansion history of the universe from when it was matter-dominated, through to the current epoch of dark-energy domination \citep{Riess1998, Perlmutter1999}. This Hubble diagram may be compared to the theoretical prediction of a given homogeneous model, in order to determine $\Omega_{\rm m}$ and the equation of state of dark energy in simple extensions of $\mathrm{\Lambda CDM}$ \citep{Scolnic2018}. If the absolute magnitudes of SN Ia are calibrated then the Hubble constant $H_0$ is also measured (see for example \citet{Shah2021} for a review). A standard data set often used in cosmological analyses is Pantheon \citep{Scolnic2018}, which combines observations from multiple surveys, and we use Pantheon in this note. 
\par
The accuracy of cosmological parameters determined from SN Ia depends on reliable standardisation and bias corrections, which is done in Pantheon via a distance modulus estimator \citep{Tripp1999} : 
\begin{equation}
\label{eq:tripp}
\mu = m_B - M_B + \alpha x_1 - \beta c + \Delta_{\rm M} + \Delta_{\rm B} \; ,
\end{equation}
where the observables are $m_B$, the log of the flux normalisation of the SN Ia lightcurve; $x_1$, a stretch parameter determined by the duration of the light curve; and $c$, the deviation of the $B-V$ colour from the mean. $M_B$ is the absolute magnitude of a fiducial mean SN Ia light curve. $\Delta_{\rm M}$ accounts for environmental effects in the host galaxy, and $\Delta_{\rm B}$ is a Malmquist bias correction. 
\par
Weak gravitational lensing, which causes non-Gaussian fluctuations in the magnitudes of SN Ia, is usually treated as a source of scatter. SN Ia seen on over-dense lines of sight (compared to a homogeneous universe of the same average matter density) will be magnified, and those on under-dense ones de-magnified. Although the distribution of lensing is \textit{not} Gaussian, a rough guide to its size is the r.m.s. scatter $\sigma_{\rm lens} = (0.06 \pm 0.017) (d_C(z)/d_C(z=1))^{3/2}$ mag, where $d_C(z)$ is the comoving distance to redshift $z$ (\citet{Shah2022}, hereafter S22).
\par 
There are two principal effects of lensing on cosmological parameter estimation. If treated as a source of noise, the accuracy of high-z SN Ia data is degraded as $\sigma_{\rm lens}$ approaches the intrinsic scatter of SN Ia \citep{Holz2005}. Secondly, if SN Ia are preferentially selected depending on whether they are magnified or not, a bias is introduced. While this bias may in principle be simulated and incorporated into $\Delta_{\rm B}$ in Equation \ref{eq:tripp} (see \citet{Kessler2017} for details), it requires an estimation of the lensing probability distribution which was not available to Pantheon at the time. A further assumption is that the \textit{actual} selection process of SN Ia is modelled by the simulation and that the numbers of SN Ia suffice to converge to the mean. 
\par 
In this note, we have two goals. Firstly, we compare values of the weak lensing convergence estimator proposed in S22 between the lines of sight to Pantheon SN Ia and randomly chosen ones. This is to statistically test if residual bias exists, and to what equivalent magnitude. Secondly, we use de-lensed magnitudes to evaluate the effect this may have on cosmological parameters.

\section{Modifying the Tripp estimator for weak lensing}
\label{sec:snwl}
Working to first order in weak lensing convergence $\kappa$, the change in magnitude $m$ is $\delta m  = -(5/\log{10}) \kappa + \mathcal{O}(\kappa^2, \gamma^2)$ where $\gamma$ is the shear. We may write the magnification of supernovae $i$ as the sum of the contribution from $N_i$ individual lenses as 
\begin{equation}
\label{eq:deltamprime}
    \Delta m_i^{'} = \sum_{j=1}^{N_i} \delta m_{ij}(\mathbf{\lambda}) \;.
\end{equation}
$\Delta m_i^{'}$ is defined as the magnification relative to the \say{empty beam} value\footnote{This is the magnification relative to a homogeneous, under-dense background matter density. The virial haloes are superimposed on this.}, so $\Delta m_i^{'}<0$ always. As detailed in S22, by taking lensing as due to dark matter haloes surrounding foreground galaxies, the parameters determining $\delta m_{ij}$ are $\mathbf{\lambda} = \{ M_r, z_s, z_d, \theta, \Gamma, \beta \}$ are the r-band absolute magnitude of the galaxy, the spectroscopic redshift of the source, photometric redshift of the lens and its angular distance to the line of sight, mass-to-light ratio $\Gamma$ and halo radial matter density slope $\beta$ respectively. Again to first order in the convergence, flux conservation may be used to express the estimated magnification of a given SN Ia as
\begin{equation}
\label{eq:deltam}
    \Delta_{\rm lens} = \Delta m_i^{'} - \langle \Delta m' (z_i) \rangle  \;,
\end{equation}
The average in the second term is over random lines of sight to the same redshift as the source. The distribution of $\Delta_{\rm lens}$ is skewed : large numbers of slightly de-magnified SN Ia will be balanced by smaller numbers of magnified SN Ia (for a quantification of skew, see \citet{Kainulainen2009}). To first order, averaging over sources is equivalent to averaging over area \citep{Kaiser2016}.
\par
By correlating the lensing estimator $\Delta_{\rm lens}$ to the Hubble diagram residual $\mu_{\rm res} = \mu - \mu_{\rm model}$ where $\mu_{\rm model}(z)$ is the distance modulus of a best-fit homogeneous cosmology, Bayesian estimates for the mass-to-light ratio $\Gamma$ and radial slope $\beta$ may be found. S22 found $\beta = 1.8 \pm 0.3$ (slightly shallower than the NFW profile \citep{Navarro1996} which has $\beta = 2$), and $\Gamma = 197 ^{+64}_{-80} \, h \, M_{\odot}/L_{r, \odot}$ where $65\%$ credibility intervals are indicated. Strictly speaking we should marginalise over the full posterior pdf $p(\Gamma, \beta)$, but for the purposes of this note it suffices to use the mean values. These were determined using Pantheon and foreground galaxy data from the Sloan Digitial Sky Survey (SDSS) \citep{Eisenstein2011}. We therefore modify the Tripp estimator as 
\begin{equation}
\label{eq:trippnew}
\mu = m_B - M_B + \alpha x_1 - \beta c  + \Delta_{\rm M} + \Delta_{\rm B} - \eta \Delta_{\rm lens}\; ,
\end{equation}
where $\eta$ is a parameter to adjust for the magnitude limitation of the SN Ia and galaxy data ($\eta = 1$ for volume-limited surveys). It was found in S22 that $\eta \simeq 1$ is adequate for the Pantheon and SDSS combination. Here $\Delta_{\rm lens}$ is a line-of-sight \say{environmental} variable akin to $\Delta_{\rm M}$. The minus sign is for de-lensing, analogous to the de-reddening term $\beta c$.  
\par
For the second term in Equation (\ref{eq:deltam}) we generate 10,000 lines of sight at random within the SDSS footprint, terminating in a redshift that has been randomly selected from Pantheon (that is, the random lines of sight shuffle only the sky positions of SN Ia and not their redshift distribution). We use the same selection criteria as S22 to remove heavily masked fields (but do not select fields on whether a host has been identified or not; most of our random LOS will intentionally not lie near any putative host galaxy). 
\par
For Pantheon, we calculate $\Delta_{\rm lens}$ for the 727 SN Ia lying within the SDSS footprint between $0.05 < z< 1.0$ (SDSS is not deep enough to reliably estimate lensing for SN Ia with $z>1$). $\Delta_{\rm lens}$ varies between $-0.19 $ and $0.09$ and has a skew of $-3.0$. The standard deviation increases with the comoving distance to the source per the formula given above.

\section{Are SN Ia lines of sight special?}
\label{sec:snspecial}
SN Ia candidates far outnumber those confirmed, with the main constraint being the observing time on spectroscopic platforms. Selection occurs both with the detection of candidates (\textit{detection efficiency}) and spectroscopic confirmation (\textit{selection efficiency}). It is argued in \citet{Scolnic2018} that both selections are well-characterised by an estimated probability $f(m_r) \in (0,1)$ which depends solely on the r-band magnitude $m_r$ of the SN Ia (see section 3.3 of \citet{Scolnic2018}). The observed population does indeed show a drift with redshift : supernovae from the high-z SNLS subset are bluer and more stretched than the Low-z subset, amounting to $\sim 0.2$ magnitudes brighter before standardisation. The $\Delta_{\rm B}$ term in Equation (\ref{eq:tripp}) corrects the light curve parameters for the consequent Malmquist bias. It is calculated by simulations convolving a model for the scatter of SN Ia parameters with $f(m_r)$ according to the method outlined in \citet{Kessler2017}. For Pantheon, a $\sigma_{\rm lens}$ term was included in the covariance matrix (see Eqn. \ref{eq:pancov} below) but not in the bias correction calculation.\footnote{More recent datasets such as the Dark Energy Survey \citep{Smith2020} and PantheonPlus \citep{Scolnic2022} do incorporate lensing probability distributions derived from N-body simulations into their bias calculations \citep{Kessler2019, Brout2022}. In this case, $\Delta_{\rm B}$ should be re-calculated before de-lensing the distances as done in Eqn. \ref{eq:trippnew}.}
\par 
Can magnitude be the sole selection criteria for SN Ia? \citet{Weinberg1976} noted that if background SN Ia are obscured by foreground galaxies, sources are de-magnified on average towards the \say{empty beam} value. In our model this would be the luminosity distance in a homogeneous universe where all matter associated with virialised halos has been removed. He argued a \say{radius of avoidance} of $R \sim 10$kpc around a foreground galaxy is sufficient to trigger this effect (a quantitative prescription in terms of survival probability of a light ray is given in \citet{Kainulainen2011}). SN Ia may not need to be strictly obscured : a desire to avoid blending or fibre collisions with foreground galaxies in spectroscopic selection would create an equivalent effect. 
\par
One could then argue for bias in either direction. Also, Pantheon combines multiple surveys each of which has their own selection characteristics. It is therefore worthwhile to compare the distribution of $\Delta m'$ for the lines of sight in Pantheon to the random set, which are the two terms in Equation (\ref{eq:deltam}). 
\par
\begin{figure}
    \centering
    \includegraphics[width=0.9\columnwidth]{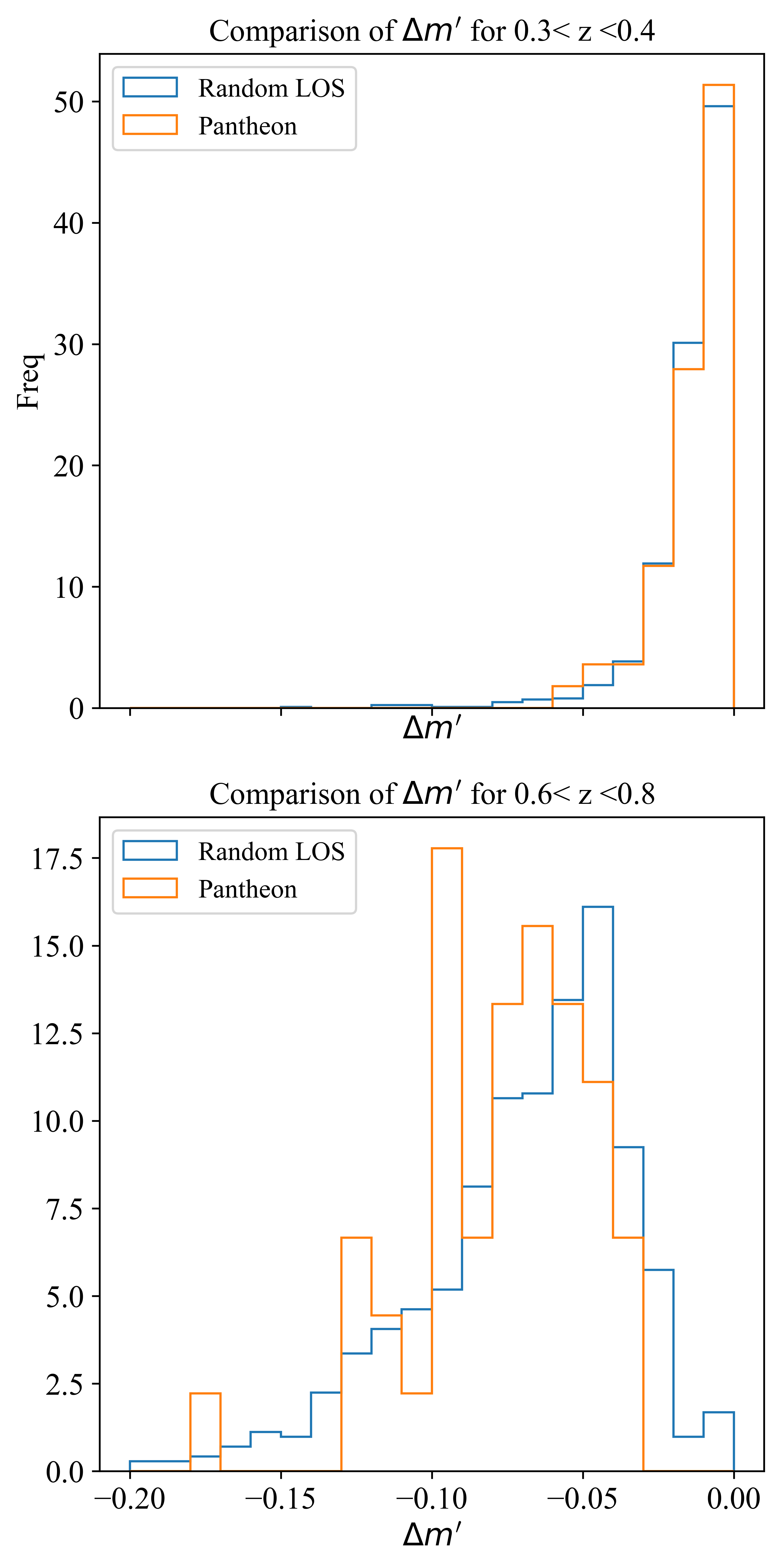}
    \caption{Comparison of the distribution of $\Delta m'$ (Equation \ref{eq:deltamprime}) for two selected redshift bins. The frequency from the random catalog has been normalised to match the numbers in Pantheon. \textit{Top:} the lines of sight to Pantheon SN Ia at $z \in (0.3, 0.4)$ are consistent with the random sample. \textit{Bottom:} Conversely high-z Pantheon SN Ia lie somewhat preferentially on over-dense lines of sight. Pantheon also lacks well-magnified $\Delta m < -0.15$ SN Ia, although this may be due to the moderate number of data points.}
    \label{fig:KStest}
\end{figure}

 
We perform a two-tailed Kolmogorov-Smirnov test in broad redshift bins (chosen to have $\sim 70$ Pantheon SN Ia each). The results are summarised in Table \ref{tab:pvals}. While there is some statistical inconsistency in the two low redshift buckets, indicating a preference to find SN Ia in clusters, these bins will not be significant for cosmological fits. In general, Pantheon lines of sight are somewhat over-dense compared to the random set, but not at large significance. The exceptions are the two bins $0.25 < z< 0.3$ and $0.6 < z < 0.8$, which interestingly are close to the magnitude limits of the main surveys in Pantheon (SDSS, SNLS and Pan-Starrs; see Figure 10 of \citet{Scolnic2018}). These are over-dense at a significance of around $2\sigma$; supernovae in these bins seem not to be on typical lines of sight. Figure \ref{fig:KStest} illustrates the distribution for two bins.
\par

\par

\begin{table}
\begin{tabular}{l l|l|c c|c c|c}
    \bfseries $z_{\rm min}$ & \bfseries $z_{\rm max}$ & \bfseries N & \bfseries Random $\langle \Delta m' \rangle$  & \bfseries Pantheon $\langle \Delta m' \rangle$ & \bfseries p-value
    \begin{filecontents*}{KStest3.csv}
zmin,zmax,rlen,ravg,rmin,rstd,rskew,plen,pavg,pmin,pstd,pskew,d,pval
0.05,0.1,482,-0.0003,-0.0129,0.001,-7.3662,48,-0.0008,-0.0119,0.002,-3.9512,0.2734,0.0022
0.1,0.15,817,-0.0012,-0.0556,0.0031,-8.9743,80,-0.0031,-0.0588,0.0086,-5.2219,0.1971,0.0059
0.15,0.2,1057,-0.0028,-0.0306,0.0041,-3.2004,110,-0.0039,-0.0776,0.0086,-6.4462,0.0922,0.3429
0.2,0.25,1047,-0.0053,-0.0675,0.0067,-3.7085,108,-0.0056,-0.0587,0.0073,-4.2622,0.0701,0.6921
0.25,0.3,981,-0.0081,-0.1579,0.0107,-5.9995,90,-0.0098,-0.0611,0.0098,-2.4704,0.1538,0.036
0.3,0.4,1276,-0.0143,-0.1477,0.0141,-3.4209,114,-0.0136,-0.0569,0.0116,-1.8616,0.0466,0.9685
0.4,0.5,628,-0.0269,-0.2086,0.0202,-2.7269,52,-0.0336,-0.1594,0.0292,-2.4544,0.1213,0.4433
0.5,0.6,528,-0.0479,-0.2152,0.0299,-1.7094,37,-0.0482,-0.1702,0.0308,-2.0331,0.0595,0.9989
0.6,0.8,728,-0.073,-0.3424,0.0435,-2.0961,48,-0.0825,-0.2611,0.0433,-2.1514,0.2205,0.0211
0.8,1.0,538,-0.0939,-0.2955,0.0454,-1.0599,39,-0.0905,-0.2073,0.0473,-0.5208,0.0906,0.8977
\end{filecontents*}
\csvreader[head to column names]{KStest3.csv}{}
    {\\\hline \zmin & \zmax & \plen & \ravg & \pavg & \pval } 
\end{tabular}

\caption{Comparison between the empty-beam average $\langle \Delta m' \rangle$ computed on random lines of sight within the SDSS footprint, and the lines of sight to Pantheon SN Ia. There is a general trend for Pantheon SN Ia to appear more magnified than the random sample. The probability-to-exceed of a two-tailed Kolmogorov-Smirnov test is given in the last column. While the distributions are compatible for many buckets, in particular for $z \in (0.25,0.3)$ and $z \in (0.6,0.8)$ the distributions are distinct at moderate significance. These buckets roughly correspond to the magnitude limits of the SDSS, Pan-Starrs and SNLS SN Ia surveys respectively. }
\label{tab:pvals}
\end{table}

\section{Cosmological parameters}
We ran a fit for cosmological parameters on the 727 Pantheon SN Ia for which we have calculated $\Delta_{\rm lens}$. We use the Pantheon covariance matrix $C = C_{\rm stat} + C_{\rm sys}$ (Equation 6 of \citet{Scolnic2018}) for our likelihood where $C_{\rm sys}$ is derived from the calibration of SN Ia lightcurves, and
\begin{equation}
\label{eq:pancov}
    C^{\rm stat}_{ii} = \sigma_{N}^{2} + \sigma_{\rm Mass}^{2}+ \sigma_{\rm v-z}^{2}+ \sigma_{\rm lens}^{2} + \sigma_{\rm int}^{2} + \sigma_{\rm Bias}^{2} \;,
\end{equation}
is a collection of diagonal terms which in particular assume lensing is a Gaussian random scatter term of size given by $\sigma_{\rm lens} = 0.055z$ \citep{Jonsson2010}. The Pantheon likelihood is $\mathcal{L} = \exp(-\chi^2 /2)$ where 
\begin{equation}
    \chi^2 = \mathbf{\mu_{\rm res}^{T}  C^{-1}  \mu_{\rm res}} \;.
\end{equation}
Here $\mu_{\rm res} = \mu - \mu_{\rm model}$ where $\mu$ is as published in \citet{Scolnic2018} (baseline), or modified by Equation (\ref{eq:trippnew}) (de-lensed). For the de-lensed case, we adjust the Pantheon covariance diagonal elements to remove the stated lensing variance: $C' = C - (0.055z)^2$. The additional variance due to the uncertainty in our lensing estimator is small compared to other variance terms and we have found adding it does not affect the results. The standard FRLW cosmological formulae are used for $\mu_{\rm model}$. Our baseline values are consistent with, but somewhat different to \citet{Scolnic2018} due to the smaller numbers of SN Ia used. 
\par
Cosmological parameters are often fit using the minimum-$\chi^2$ method (see for example Section 6 of \citet{Scolnic2018}). Seen from a Bayesian perspective, this is equivalent to a maximum-likelihood if the likelihood is Gaussian, of known covariance and a uniform prior is taken for the cosmological parameters. In other words, the \textit{mode} of the posterior distribution is found. However as lensing is a skewed distribution, a likelihood incorporating its effect will be non-Gaussian. It is then possible that de-lensing may result in larger differences in maximum likelihood values compared to the mean values. One can understand the effect on the dark energy equation of state parameter $w$ in the $w$CDM model as follows. If $w<-1$, SN Ia appear dimmer compared to $\Lambda$CDM. However, as the maximum likelihood value for lensing also results in a dimmer SN Ia, de-lensing may be expected to increase the maximum likelihood of $w$. \citet{Amendola2010} and \citet{Holz2005} investigated this using simulations, and state $w$ is biased lower by $0.1$ if the likelihood is falsely assumed to be Gaussian. However a similar analysis by \citet{Sarkar2008} found minimal bias. Unmodelled systematics in the observing strategy as described in Section \ref{sec:snspecial} may also contribute to skew effects.
\par 
We are therefore motivated to compare both the modes and mean of the posterior between the baseline and de-lensed cases given uniform priors of $\Omega_{\rm m} \in (0.2, 0.4)$ and $w \in (-0.5, -1.5)$. Our cosmological parameter shifts are unaffected by whether we fix the nuisance parameter $M=-19.425$ (which is equivalent to $H_0 = 67.4 \;\;\SI{}{\kilo\meter\per\second\per\mega\pc}$) or marginalise over it. We also assume that the nuisance parameters $\alpha$ and $\beta$ are unaffected by lensing. The Polychord package \citep{Handley2015} was used to generate the chains.

\section{Results}
\begin{figure}
    \centering
    \includegraphics[width=\columnwidth]{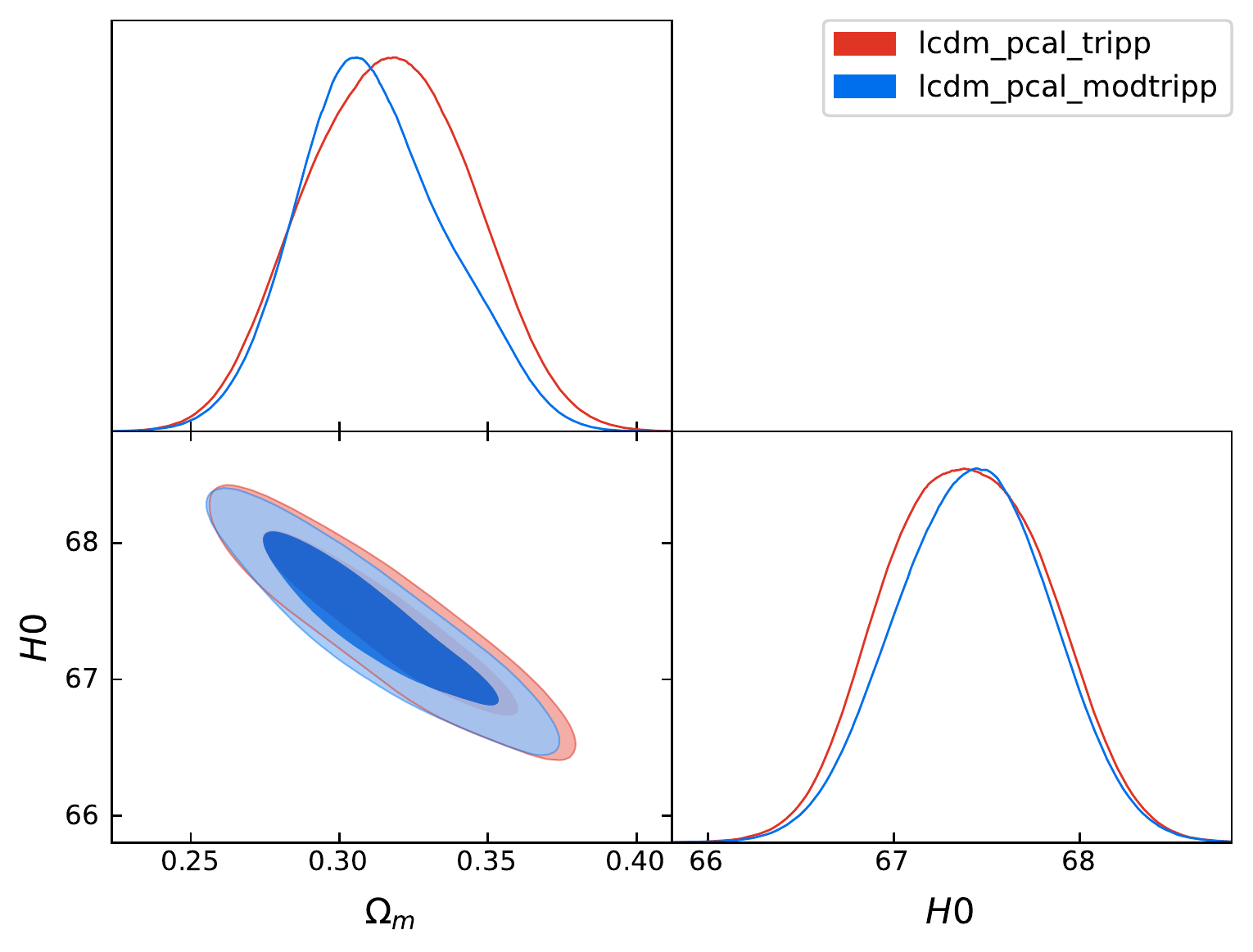}
    \caption{Comparison of $\Omega_M$ and $H_0$ with a fixed SN Ia calibration of $M = -19.425$ in $\Lambda$CDM. The posterior of the original Tripp estimator (Equation (\ref{eq:tripp}) is shown in red, and the de-lensed estimator (Equation (\ref{eq:trippnew}) is blue.}
    \label{fig:triangle}
\end{figure}

For $\Lambda$CDM, we find the mean $\Omega_{\rm m} = 0.317 \pm 0.027$ (baseline) and $\Omega_{\rm m} = 0.312 \pm 0.025$ (de-lensed). For Flat $w$CDM, we find $\Omega_{\rm m} = 0.332 \pm 0.049$ and $w = -1.16 \pm 0.16$ (baseline), whereas $\Omega_{\rm m} = 0.316 \pm 0.048$ and $w = -1.12 \pm 0.15$ for the de-lensed case. The posteriors for $\Lambda$CDM for are shown in Figure \ref{fig:triangle}. $\Delta \chi^2 \sim -15$ in both $\Lambda$CDM and $w$CDM for the de-lensed data.

\par 
The differences of $\Delta \Omega_{\rm m} = -0.005 \pm 0.002$ ($\Lambda$CDM) and $(\Delta \Omega_{\rm m}, \Delta w) = (-0.016 \pm 0.008 , +0.04 \pm 0.02)$ ($w$CDM) are low compared to the cosmological uncertainty 
(of the overall Pantheon sample). We have computed the errors on the shifts by sampling from the posteriors for the model parameters and photometric redshifts. Our statistical test of the preceding paragraph isolated the significance of the lensing correction independent of the background cosmological uncertainty. 
\par
The maximum likelihood value changes for $\Lambda$CDM by $\Delta \Omega_m = -0.006$, and for $w$CDM the change is $(\Delta \Omega_{\rm m}, \Delta w) = (-0.017, +0.04)$. We emphasize we have computed the change for realised Pantheon sample, rather than the average of a simulated ensemble. 

\section{Conclusion}
Our answer to the question \say{are SN Ia biased probes of cosmological parameters?} is then a qualified \say{no} for Pantheon. We have found moderate evidence of an uncorrected bias to select magnified SN Ia in two buckets around the magnitude limits of the main contributing surveys. However, the change in cosmological parameters caused by de-lensing SN Ia magnitudes is not significant compared to their uncertainty. We find no evidence that a \say{zone of avoidance} around foreground galaxies has selected SN Ia on under-dense lines of sight, although it may be that some offset between the two effects is present. 
\par 
Our test can be performed on any future survey to prospectively or retrospectively check for bias, and uses only observational data with no need for simulations of the lensing pdf. It will be particularly well-suited to high-z surveys such as the Rubin LSST and the Roman Space Telescope, due to the larger effect of lensing and also by being paired with a galaxy catalog assembled from the same platform. It may also be necessary to consider $\mathcal{O}(\kappa^2)$ effects at higher redshifts, which can introduce further bias \citep{Kaiser2016}.

\begin{figure}
    \centering
    \includegraphics[width=\columnwidth]{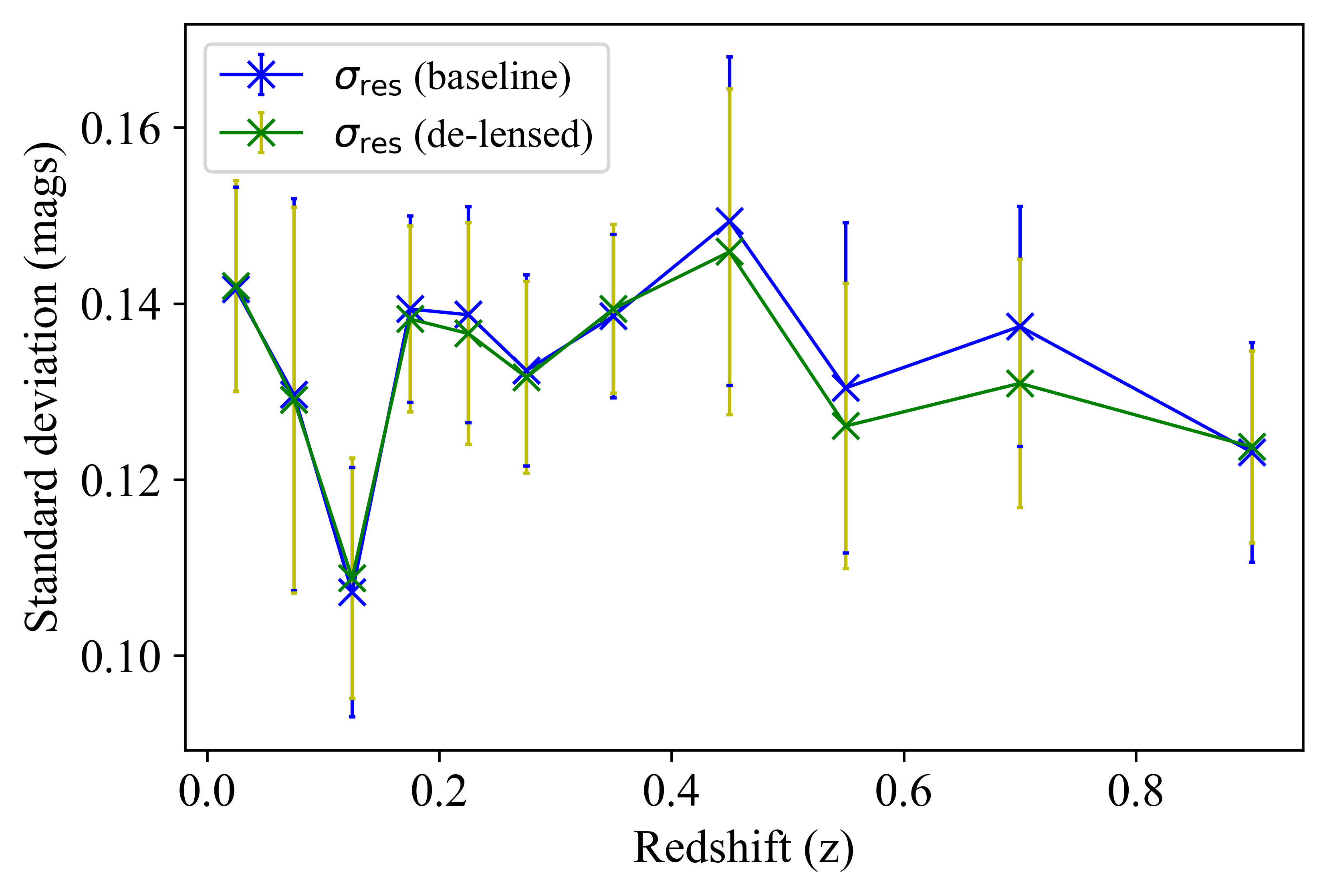}
    \caption{Standard deviation of residuals of a Hubble diagram fit. The de-lensed residuals are lower by $\sim 0.005$ for $z>0.4$, with the exception of the last bucket (see S22 for a discussion of why these SN Ia -- observed by the Hubble Space Telescope in a narrow field -- are different). Error bars have been constructed by bootstrap resampling.}
    \label{fig:residuals}
\end{figure}

\section*{Acknowledgements}
OL and PL acknowledge support from an STFC Consolidated Grant ST/R000476/1, and PL acknowledges STFC Consolidated Grant ST/T000473/1.

\section*{Data Availability}
Upon publication, a file containing computed $\Delta_{\rm lens}$ for Pantheon will be made available at \url{https://github.com/paulshah/SNLensing}.
 



\bibliographystyle{mnras}
\bibliography{SnMagCitations2} 


\bsp	
\label{lastpage}
\end{document}